\documentclass[aps,floatfix,superscriptaddress,reprint,10pt,pra]{revtex4-1}
\usepackage{bbm}
\usepackage{bm}
\usepackage{amsmath}
\usepackage{amssymb}
\usepackage{empheq}
\usepackage{graphicx}
\usepackage{mathrsfs}
\usepackage{amsfonts}
\usepackage{amsthm}
\usepackage{color}
\usepackage{bigints}
\usepackage{txfonts}
\usepackage{hyperref}
\usepackage{color}
\usepackage{appendix}
\usepackage{multirow}
\usepackage{makecell}

\hypersetup{
	unicode=false,               pdftoolbar=true,             pdfmenubar=true,             pdffitwindow=false,          pdfstartview={FitH},         pdftitle={My title},         pdfauthor={Author},          pdfsubject={Subject},        pdfcreator={Creator},        pdfproducer={Producer},      pdfkeywords={keyword1} {key2} {key3},      pdfnewwindow=true,           colorlinks=false,            linkcolor=red,               citecolor=green,             filecolor=magenta,           urlcolor=cyan           }
\makeatletter \tolerance = 10000 \tolerance = 10000
\makeatother

\begin{document}
 
\title{ Cross-Layer Anomalous Hall Transport driven by N\'eel-Vector rotating in the Altermagnet candidate V$_2$Te$_2$O}

\author{Yanan Pan}
\thanks{These two authors contributed equally to this work.} 
\affiliation{Department of Physics, Hefei University of Technology, Hefei, Anhui 230601, China}

\author{W. Z. Zhuo}
\thanks{These two authors contributed equally to this work.} 
\affiliation{School of Optoelectronic Engineering, Guangdong Polytechnic Normal University, Guangzhou 510665, China}

\author{Pan Gao}
\affiliation{Department of Physics, Hefei University of Technology, Hefei, Anhui 230601, China}

\author{Ziyu Zhou}
\affiliation{Department of Physics, Hefei University of Technology, Hefei, Anhui 230601, China}

\author{Junqing Xu}
\affiliation{Department of Physics, Hefei University of Technology, Hefei, Anhui 230601, China}

\author{Lijie Shao}
\affiliation{Department of Physics, Hefei University of Technology, Hefei, Anhui 230601, China}

\author{Damin Meng}
\affiliation{Department of Physics, Hefei University of Technology, Hefei, Anhui 230601, China}

\author{Weiwei Chen}
\thanks{Contact author. E-mail: chenweiwei@hfut.edu.cn}
\affiliation{Department of Physics, Hefei University of Technology, Hefei, Anhui 230601, China}

\author{Zhongjun Li}
\thanks{Contact author. E-mail: zjli@hfut.edu.cn}
\affiliation{Department of Physics, Hefei University of Technology, Hefei, Anhui 230601, China}

\author{Ye Yang}
\thanks{Contact author. E-mail: yangye@hfut.edu.cn}
\affiliation{Department of Physics, Hefei University of Technology, Hefei, Anhui 230601, China}

\date{\today}

\begin{abstract}
In van der Waals (vdW) materials, weak interlayer coupling generally suppresses vertical dispersion, reinforcing the conventional paradigm that in-plane transport dominates over cross-layer channels. Here, using first-principles calculations and magnetic symmetry analyses, we uncover a giant, symmetry-unlocked cross-layer anomalous Hall conductivity (AHC) in the vdW altermagnet V$_2$Te$_2$O. In the magnetic ground state with N\'eel vector $\bm{N}\parallel z$, horizontal mirror symmetry protects a spin-polarized nodal chain near the Fermi level and strictly enforces zero anomalous Hall response. Tilting the N\'eel vector explicitly breaks this mirror protection, allowing spin-orbit coupling to gap the nodal chain and activate a sharp cross-layer Hall response. When the N\'eel vector is rotated into the in-plane configuration ($\bm{N}\parallel x$), cross-layer orbital hybridization generates intensive Berry curvature hotspots, boosting the cross-layer component of AHC $\sigma_{yz}$ to approximately 255 S/cm, which exceeds in-plane component $\sigma_{xy}$ by nearly two orders of magnitude. Furthermore, varying the azimuthal angle systematically redistributes the anomalous Hall response, enabling full directional control of transverse transport. Our findings demonstrate a highly sensitive cross-layer anomalous Hall switch activated by low-barrier spin canting, offering promising avenues for directional tensor selection and low-power multi-axial vdW spintronics. 

\end{abstract}

\maketitle

\textit{Introduction.---}
Generating an out of plane transverse Hall current perpendicular to the van der Waals (vdW) basal plane in response to an in plane electric field provides an exceptional spintronic transport geometry. By spatially decoupling the out-of-plane transverse signal from the large in-plane driving current, this cross-layer configuration eliminates background Ohmic voltages and offers a high signal-to-noise ratio~\cite{Tiwari2021nc}. However, because weak interlayer electronic coupling inherently suppresses vertical dispersion, a long-standing paradigm assumes that transverse Hall transport in vdW material is predominantly confined within basal planes~\cite{Deng2020science,Tan2021nl,Liu2024nl}. Consequently, in-plane anomalous Hall conductivity (AHC, $\sigma_{xy}$) dominates, whereas cross-layer AHC components ($\sigma_{yz}$ and $\sigma_{zx}$) are generally suppressed. Overriding this intrinsic anisotropy to achieve a dominant, topologically driven cross-layer AHC remains a challenging goal.

Magnetic ordering plays a fundamental role in generating the AHC by breaking time-reversal symmetry and allowing nonzero Berry curvature in momentum space~\cite{Nagaosa2010,Xiao2010,Fakhredine2023,Schilberth2023}. Historically, research focused primarily on ferromagnets, where net magnetization directly drives the Hall voltage \cite{Lin2019prb,Deng2022nl}. The field subsequently expanded to noncollinear antiferromagnets, which showed that large AHC can emerge even without a net magnetic moment due to momentum-space Berry curvature. Studies on noncollinear antiferromagnets Mn$_3X$ ($X=$ Ir, Sn, Ge, and Pt) established that compensated magnetic order can still support a finite Hall vector when allowed by the magnetic space group~\cite{Chen2014,Zhang2017,Nakatsuji2015,Nayak2016,Kiyohara2016,GuoWang2017}. Recently, altermagnets have extended this principle to compensated collinear maagnetic systems. In conventional collinear antiferromagnets, a combined $T S$ symmetry, where $S$ denotes spatial inversion $I$ or translation $\tau$ and $T$ denotes time reversal, generally enforces band degeneracy and Berry-curvature cancellation. Altermagnets evade this constraint because their opposite-spin sublattices are connected by crystal rotations rather than space inversion or translation, yielding momentum-dependent spin splitting~\cite{Smejkal2022PRX,Smejkal2022Landscape,Smejkal2022NRM}. First-principles studies on prototype RuO$_2$ predicted strongly anisotropic anomalous Hall, Nernst, and thermal Hall responses~\cite{Smejkal2020,Zhou2024}, although its exact ground state and altermagnetic splitting remain actively debated~\cite{Hiraishi2024,Liu2024RuO2}. Spontaneous Hall responses have been observed in MnTe and Mn$_5$Si$_3$ \cite{Betancourt2023,Reichlova2024}, while photoemission measurements have established large altermagnetic band splitting in MnTe and CrSb \cite{Krempasky2024,Reimers2024,Ding2024}. Theoretical works on CrSb and monolayer V$_2$Te$_2$O further demonstrated that N\'eel-vector reorientation modifies the Berry-curvature distribution and manipulates allowed Hall tensor components \cite{Yu2025,Liu2025V2Te2O}. Related proposals for layer Hall detection in magnetoelectric antiferromagnets, field-induced in-plane AHC in $PT$-symmetric antiferromagnets, and A-type antiferromagnetic bilayers further illustrate how magnetic symmetry permits, forbids, or redistributes Hall responses in compensated antiferromagnets \cite{Tao2024,Cao2023,Liu2025Bilayers}.

The emergence of layered altermagnetic candidate materials provides a promising platform to address the limitation of cross-layer anomalous Hall transport. Meanwhile, a recent classification of anomalous-Hall N\'eel textures has formalized the relation between N\'eel-vector orientation and the angular dependence of the Hall vector~\cite{Xiao2026}. Reorienting the N\'eel-vector acts as a subtle symmetry-breaking mechanism that changes the magnetic space group without inducing structural phase transitions~\cite{Yu2025,Liu2025V2Te2O,Xiao2026}. Crucially, tilting the N\'eel vector away from high-symmetry axes activates hybridization between out-of-plane and in-plane orbital components through spin-orbit coupling (SOC), offering a viable pathway to overcome weak interlayer dispersion in vdW systems. 

\begin{figure}
	\centering
	\includegraphics[width=1\linewidth]{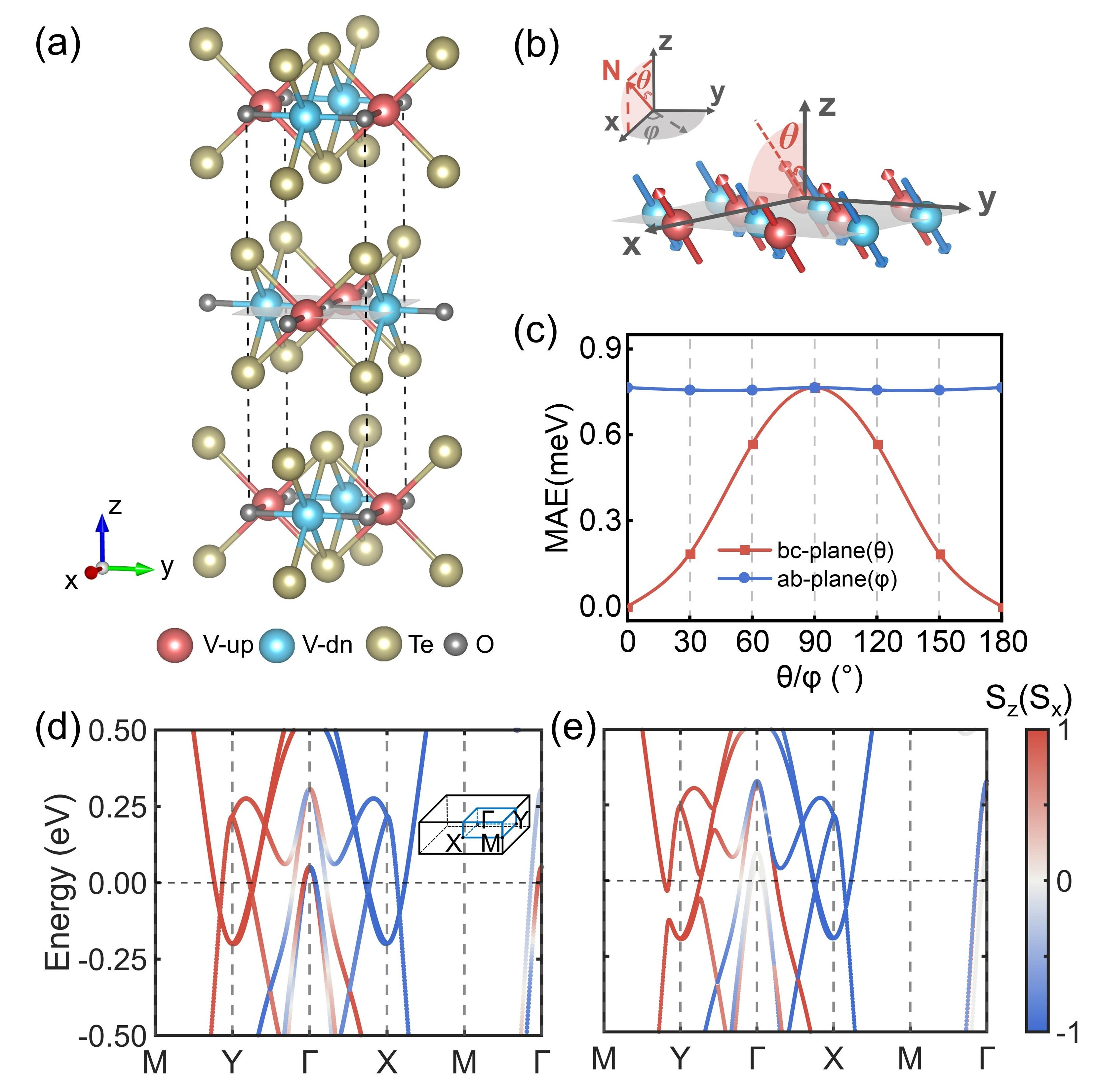}
	\caption{(a) Crystal structure of $\mathrm{V_2Te_2O}$, where the red and blue spheres represent the V atoms with opposite spin orientation, while the olive-green and gray spheres are the Te and O atoms, respectively. (b) Schematic illustration of the N\'eel-vector rotation, with polar angle $\theta$ and azimuthal angle $\varphi$. (c) MAE as a function of the N\'eel-vector orientation. (d,e) Spin-projected band structures with SOC for the N\'eel vector oriented along the $z$ and $x$ axes, respectively. The color scales represent the expectation values of $S_z$ in (d) and $S_x$ in (e), respectively. The inset in (d) shows the Brillouin zone.}
	\label{fig:fig1}
\end{figure}

In this work, using  first-principles calculations and magnetic symmetry analyses, we demonstrate a giant cross-layer AHC unlocked by N\'eel reorientation in the vdW altermagnet bulk V$_2$Te$_2$O. Magnetocrystalline anisotropy energy (MAE) calculations identify an out-of-plane magnetic ground state ($\bm{N}\parallel z$) with an exceptionally shallow rotation barrier. In this pristine configuration, horizontal $M_z$ mirror symmetry protects a spin-polarized nodal chain near the Fermi level and enforces zero AHC across all tensor components. Intriguingly, tilting the N\'eel vector slightly away from the $z$-axis breaks the $M_z$ mirror protection, allowing SOC to gap the nodal chain and activate a sharp cross-layer anomalous Hall response. When the N\'eel vector reaches the in-plane configuration ($\bm{N}\parallel x$), the cross-layer component $\sigma_{yz}$ reaches 255 S/cm, exceeding the in-plane component $\sigma_{xy}$ by nearly two orders of magnitude and overriding the conventional AHC anisotropy in vdW systems. Momentum-resolved analyses confirm that this dominant transport originates from anticrossings of the nodal chain rather than the coexisting Weyl pairs. Furthermore, tuning the azimuthal angle $\varphi$ systematically redistributes Berry curvature among all tensor components, enabling full directional control over transverse currents. Our study establishes a highly sensitive cross-layer Hall switch, opening avenues for multi-axial vdW spintronic applications.

\begin{figure}
	\centering
	\includegraphics[width=1.0\linewidth]{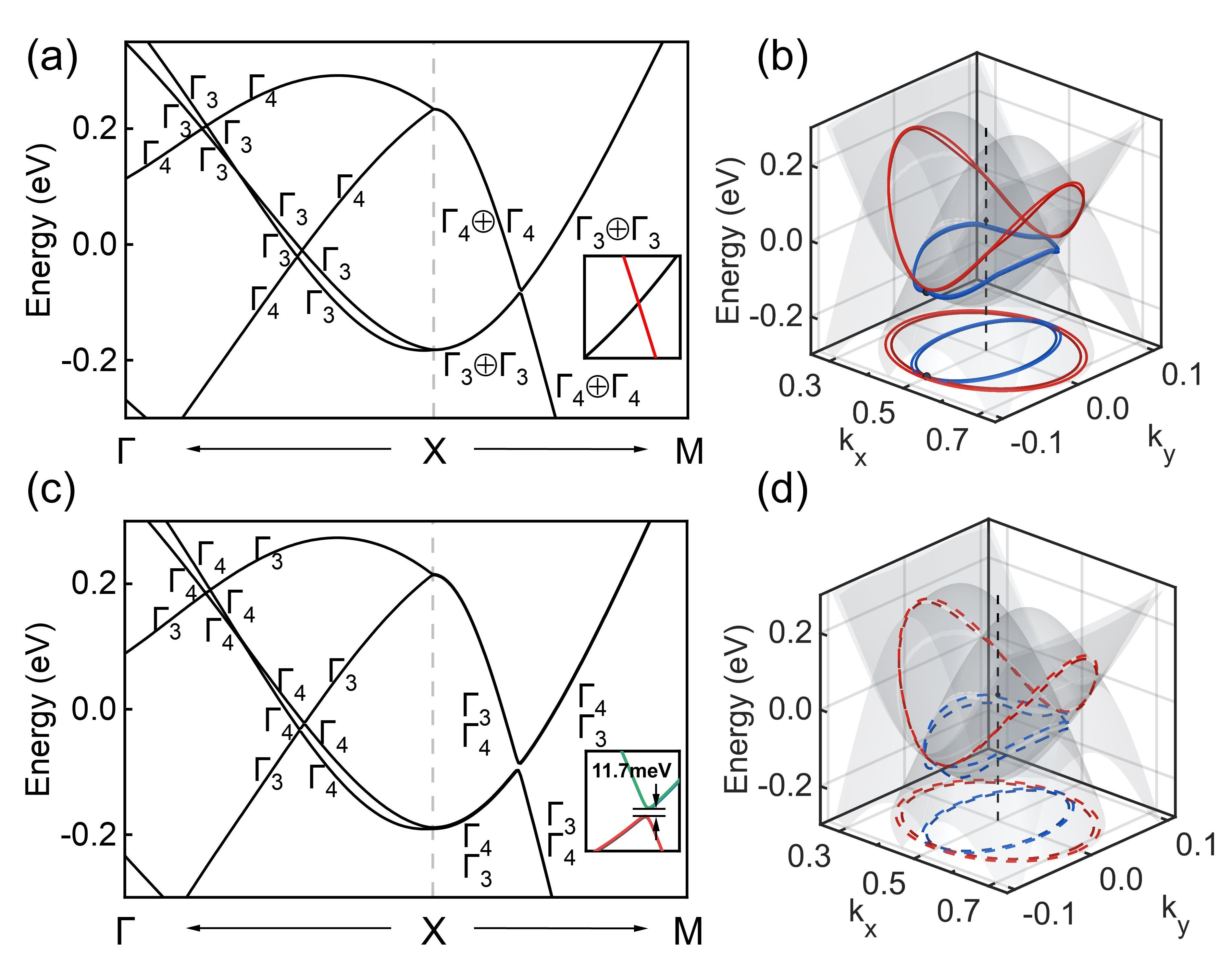}
	\caption{(a,b) Band structures near the X point for $\theta=0^\circ$ and $\theta=90^\circ$, respectively. (c,d) The Schematic illustration of the nodal ring near $X$. }
	\label{fig:fig2}
\end{figure}

\textit{Computational Methods.---}
First-principles calculations were performed using the density functional theory (DFT) framework as implemented in VASP~\cite{Kresse1996}. Exchange-correlation interactions were treated using the PBE-GGA functional~\cite{Perdew1996}, with an effective $U=3.0$~eV applied to V-$3d$ orbitals via Dudarev's GGA+$U$ method~\cite{Dudarev1998}. A plane-wave with cutoff energy of 600~eV and a $\Gamma$-centered $12\times12\times5$ $k$-point grid were used. The optimized lattice constant for bulk V$_2$Te$_2$O is $a=b=4.065$~\AA. SOC was included in all electronic-structure and transport calculations. Tight-binding Hamiltonians were constructed using Wannier90~\cite{Mostofi2014} based on V-$3d$ and Te-$5p$ projections. Dynamical stability was confirmed via phonon calculations using Phonopy~\cite{Togo2015}. Momentum-resolved Berry curvatures and AHC spectra were calculated using WannierTools~\cite{Wu2018WannierTools}. The intrinsic AHC tensor components $\sigma_{ij}$ ($ij=xy,yz,zx$) were given by ~\cite{Nagaosa2010,Xiao2010,Haldane2004,Yao2004,Wang2006},
\begin{equation}
	\sigma_{ij}=-\frac{e^2}{\hbar}\sum_n\int_{\mathrm{BZ}}\frac{d^3k}{(2\pi)^3}
	f_{n\mathbf{k}}\Omega_{n,ij}(\mathbf{k}),
\end{equation}
where $f_{n\mathbf{k}}$ is the Fermi--Dirac distribution function for band $n$ at wave vector $\mathbf{k}$, and $\Omega_{n,ij}(\mathbf{k})$ represents the Berry curvature
\begin{equation}\label{eq:Berry_curvature}
	\Omega_{n,ij}(\mathbf{k})
	=-2\hbar^2\,\mathrm{Im}\sum_{m\ne n}
	\frac{\langle n\mathbf{k}|v_i|m\mathbf{k}\rangle
		\langle m\mathbf{k}|v_j|n\mathbf{k}\rangle}
	{(\varepsilon_{m\mathbf{k}}-\varepsilon_{n\mathbf{k}})^2}.
\end{equation}
Here, $v_i=\hbar^{-1}\partial H(\mathbf{k})/\partial k_i$ denotes the velocity operator along $i$-direction, and $\varepsilon_{n\mathbf{k}}$ is the eigenvalue of state $|n\bm{k}\rangle$.

\textit{Crystal and Band Structures.---}
Bulk V$_2$Te$_2$O crystallizes in a tetragonal structure consisting of planar V$_2$O square nets sandwiched between Te sheets~\cite{Ablimit2018IC}, as illustrated in Fig.~\ref{fig:fig1}(a). Total energy comparisons confirm that the G-type AFM configuration constitutes the lowest-energy magnetic state (details in Supplemental Materials S1). Its dynamical stability is confirmed by phonon calculation presented in Supplemental Materials S2. MAE mapping in Fig.~\ref{fig:fig1}(c) reveals an out-of-plane easy axis ($\bm{N}\parallel z$) with a shallow rotation energy barrier ($<$1meV). This remarkably small barrier suggests that the N\'eel vector $\bm{N}$ can be flexibly reoriented via moderate external magnetic fields, strain, or spin-orbit torques. 

In the ground state with $\bm{N}\parallel z$, electronic bands near the Fermi level (E$_F$) exhibit pronounced spin splitting [Fig.~\ref{fig:fig1}(d)]. The electronic states near the $X$ and $Y$ points exhibit opposite spin polarizations. Along both the $M$–$Y$ and $M$-$X$ paths, four bands meet at approximately 0.08 eV below E$_F$, forming fourfold-degenerate crossing points, whereas only twofold-degenerate crossings are observed along both the $\Gamma$–$Y$ and $\Gamma$–$X$ paths. Rotating the N\'eel vector toward the $x$-axis ($\theta=90^{\circ}$) opens a substantial energy gap near $Y$, while triggering a distinct topological reconstruction near $X$ [Fig.~\ref{fig:fig1}(e)].

\begin{figure}
	\centering
	\includegraphics[width=1.0\linewidth]{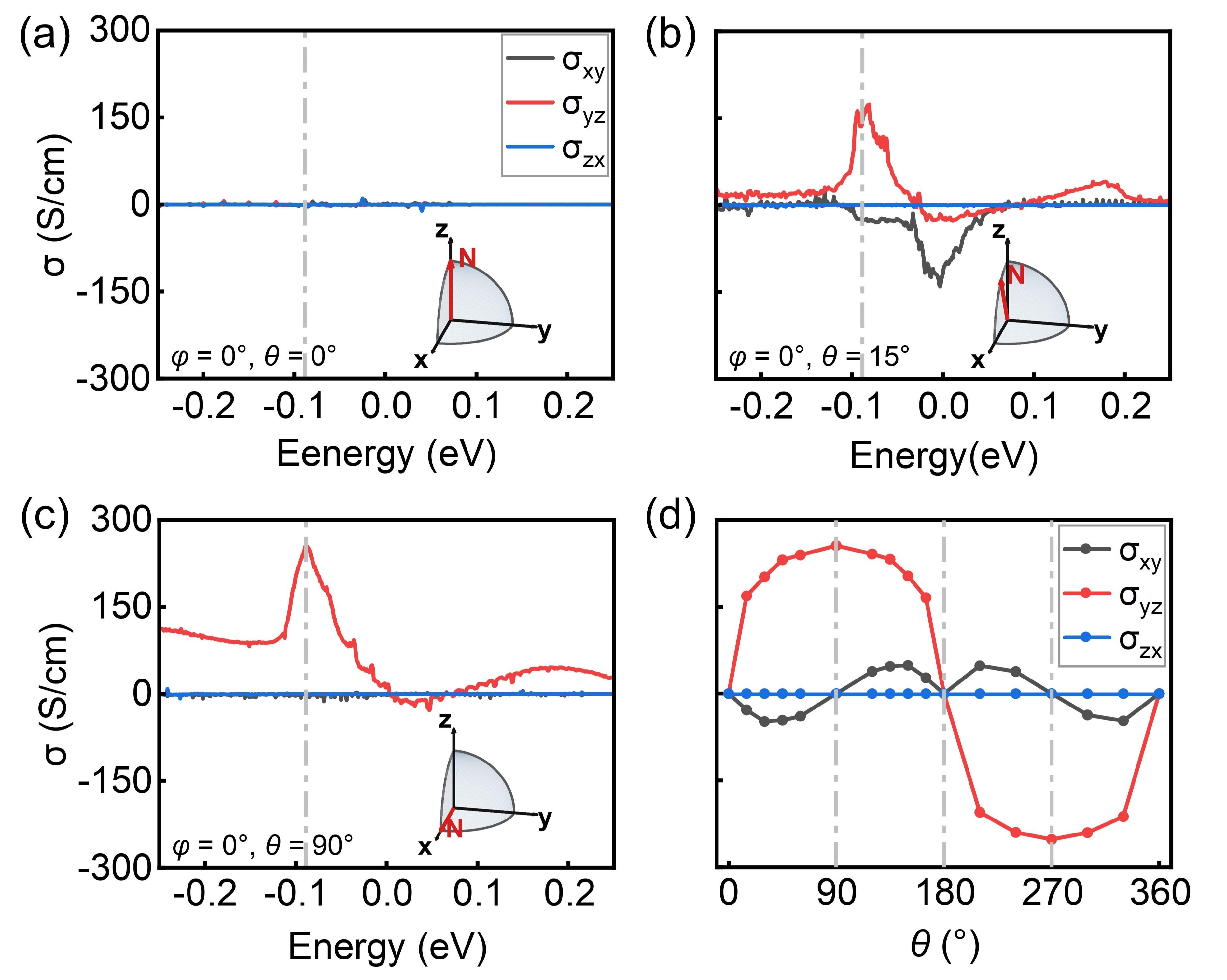}
	\caption{\textbf{N\'eel-vector-controlled anomalous Hall conductivity in $\mathrm{V_2Te_2O}$.}
		(a--c) Energy-dependent Hall-conductivity components
		$\sigma_{yz}$, $\sigma_{zx}$, and $\sigma_{xy}$ for N\'eel-vector rotations
		with $\varphi=0^\circ$ and representative polar angles
		$\theta=0^\circ$, $15^\circ$, and $90^\circ$ in the $x$--$z$ plane.
		(d) Angular dependence of the Hall-conductivity components as the N\'eel
		vector rotates within the $x$--$z$ plane. The insets indicate the $\bm{N}$ orientations in spherical coordinates.}
	\label{fig:fig4}
\end{figure}

\begin{table*}[!tbp]
	\centering
	\caption{\textbf{Symmetry constraints on anomalous Hall conductivity under  N\'eel-vector orientations.}}
	\label{tab:main_symmetry}
	\begin{tabular}{cccc}
		\hline\hline
		N\'eel-vector orientation & Magnetic group &  Generators & Forbidden AHC components \\
		\hline
		$\mathbf{N}\parallel z$ & $I4^{\prime}/mm^{\prime}m$ & $I,\ C_{2z},\ C_{2[110]},\ C_{4z}T$ & $\sigma_{xy},\ \sigma_{yz},\ \sigma_{zx}$ \\
		$0^\circ<\theta<90^\circ$ & $C2^{\prime}/m^{\prime}$ & $I,\ C_{2y}T$ & $\sigma_{zx}$ \\
		$\mathbf{N}\parallel x$ & $Im^{\prime}m^{\prime}m$ & $I,\ M_{z}T, \ C_{2y}T$ & $\sigma_{zx},\ \sigma_{xy}$ \\
		\hline\hline
	\end{tabular}
\end{table*}

\begin{figure}
	\centering
	\includegraphics[width=1.0\linewidth]{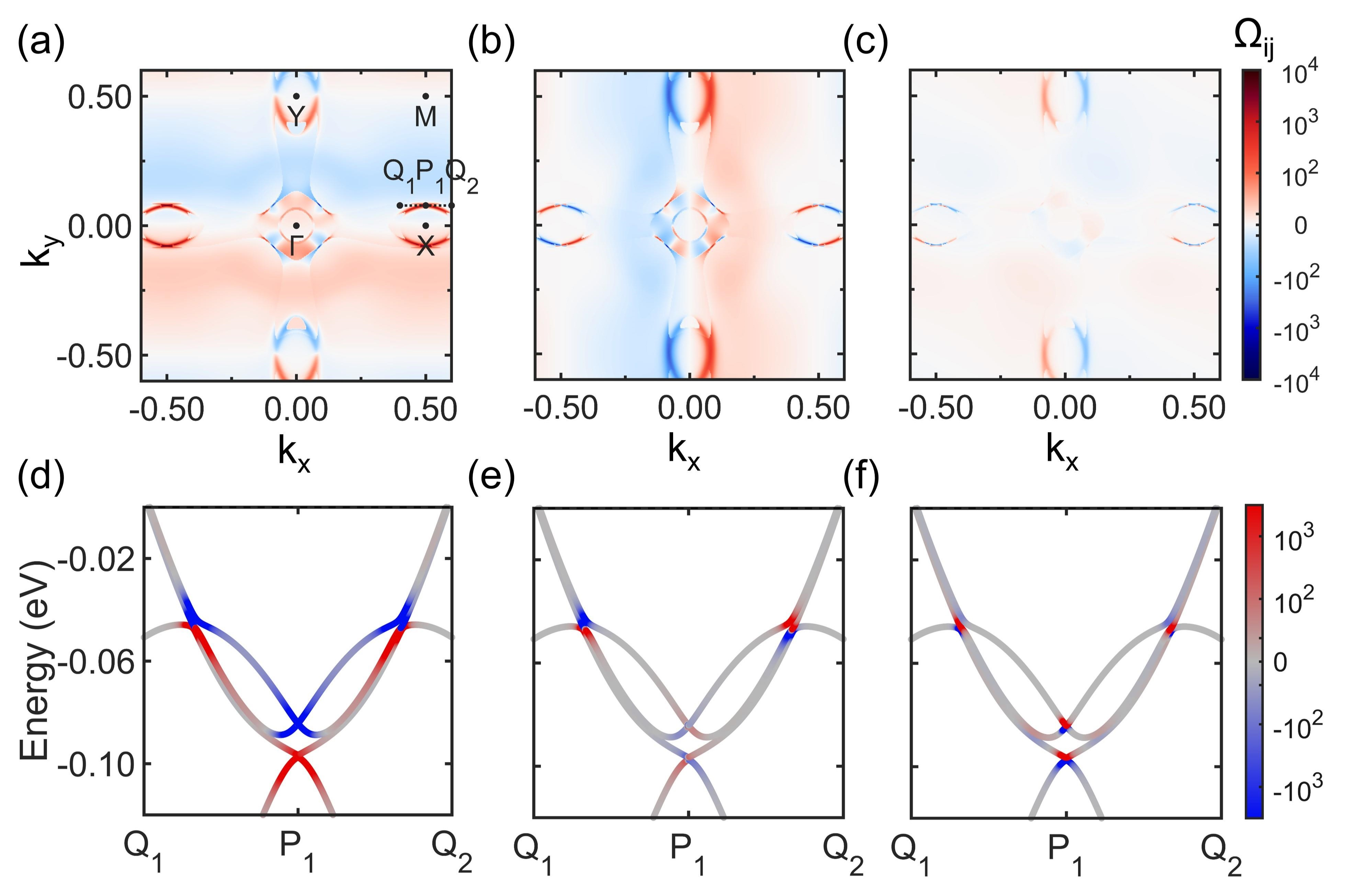}
	\caption{\textbf{Berry-curvature distributions for the tilted N\'eel-vector orientation at $\theta=90^\circ$ and $\varphi=0^\circ$.}
		(a--c) Distributions of the total Berry-curvature components
		$\Omega_{yz}$, $\Omega_{zx}$, and $\Omega_{xy}$ in the $k_z=0$ plane at the energy of the AHC peak.	The selected high-symmetry path marked in (a) is defined by $Q_1=(0.4,0.078,0)$, $P_1=(0.5,0.078,0)$, and $Q_2=(0.6,0.078,0)$. (d--f) Corresponding band-resolved Berry-curvature distributions along $Q_1$--$P_1$--$Q_2$ path. The color scale indicates both the sign and magnitude of the Berry curvature. }
	\label{fig:fig3}
\end{figure}

To uncover the topological character of these crossings, we zoom in on the dispersion near the $X$ point and present the irreducible representations of the crossing bands in Fig.~\ref{fig:fig2}. For the ground state ($\bm{N}\parallel z$), the double-degeneracy crossing along the $\Gamma$--$X$ line is protected by horizontal mirror symmetry $M_z$, with two opposite mirror eigenvalues of $+i$ and $-i$ [Fig.~\ref{fig:fig2}(a)]. Along the $X$–$M$ boundary, these four bands transform as two doubly degenerate representations ($\Gamma_{3} \oplus \Gamma_{3}$ and $\Gamma_{4} \oplus \Gamma_{4}$), intersecting bands form four nodal rings, constructing a symmetry-protected nodal chain across the Brillouin zone [Fig.~\ref{fig:fig2}(b)]. Tilting the N\'eel vector away from the $z$-axis explicitly breaks $M_z$ mirror symmetry, triggering an SOC-driven nodal-chain reconstruction. Heavy Te-5$p$ orbitals provide strong atomic SOC, mediating interlayer (V 3$d$)--(Te 5$p$)--(V 3$d$) orbital hybridization. This hybridization lifts the band degeneracy along the nodal chain, opening anticrossing gaps [Fig.~\ref{fig:fig2}(c,d)].  Additionally, an isolated pair of Weyl nodes emerges along $\Gamma$–$X$ (details in Supplemental Materials S3).

\textit{Symmetry Unlocking and Giant Cross-Layer AHC.---}
The evolution of the AHC tensor under N\'eel vector rotation is dictated by magnetic symmetry constraints on the Hall axial vector $\sigma^H=(\sigma^H_x,\sigma^H_y,\sigma^H_z)=(\sigma_{yz},\sigma_{zx},\sigma_{xy})$ (the full operation-by-operation derivation provided in Supplemental Materials S4).

In the ground state ($\bm{N}\parallel z$), V$_2$Te$_2$O belongs to the high-symmetry magnetic space group $I4'/mm'm$ generated by \{$I$, $C_{2z}$, $C_{2[110]}$, and $C_{4z}T$\}. These combined spatial and antiunitary operations force all three AHC tensor components to vanish. When $\bm{N}$ lies in the $x-z$ plane ($0^{\circ}<\theta<90^{\circ}$, $\varphi=0^{\circ}$), the symmetry lowers to $C2'/m'$, generated by \{$I$ and $C_{2y}T$\}, which forbids $\sigma_{zx}$ but unlocks finite cross-layer $\sigma_{yz}$ and in-plane $\sigma_{xy}$ components. When $\bm{N}$ is fully rotated into the in-plane $x$-axis ($\theta=90^{\circ}$, $\varphi=0^{\circ}$), the magnetic space group transforms into $Im'm'm$, generated by \{$I$, $M_{z}T$, and $C_{2y}T$\}, where symmetry forces $\sigma_{zx}=\sigma_{xy}=0$, leaving $\sigma_{yz}$ as the sole symmetry-allowed conductivity component (summarized in Table~\ref{tab:main_symmetry}).

The calculated AHC spectra [Fig.~\ref{fig:fig4}] strictly adhere to these selection rules. While all components vanish at $\theta=0^{\circ}$, a dominant cross-layer peak $\sigma_{yz}$ emerges rapidly upon tilting and reaches a giant value of $255$ S/cm near $E_F$ at $\theta=90^{\circ}$. Strikingly, $\sigma_{yz}$ exceeds the in-plane component $\sigma_{xy}$ by nearly two orders of magnitude, demonstrating extreme transport anisotropy.

To unravel the microscopic mechanism driving this giant cross-layer anomalous Hall transport, we compute the momentum-resolved Berry curvature distributions in the $k_z=0$ plane at $\theta=90^{\circ}$, $\varphi=0^{\circ}$, as presented in Fig.~\ref{fig:fig3}. Generating a dominant Berry curvature component $\Omega_{yz}$ requires finite in-plane ($v_y$) and out-of-plane ($v_z$) velocity matrix elements, alongside small energy denominators $(\varepsilon_{m\bm{k}}-\varepsilon_{n\bm{k}})^2$ near the Fermi level. Although out-of-plane dispersion is intrinsically weak in pristine vdW V$_2$Te$_2$O, the heavy Te atoms bridging the Te-V$_2$O-Te sublayers possess strong atomic SOC. Reorienting $\bm{N}$ lifts $M_z$ mirror symmetry, enabling SOC to dynamically hybridize in-plane V 3$d$ orbitals ($d_{xy}$, $d_{x^2-y^2}$) with out-of-plane orbitals (V 3$d_{xz/yz}$ and bridging Te 5$p_z$). This cross-layer orbital hybridization ($d_{\parallel}\xleftrightarrow{\text{SOC}} d_{\perp}/p_z$) creates an effective electronic channel across the vdW gap, strongly amplifying the out-of-layer velocity matrix elements $\langle m\bm{k}|v_z|n\bm{k}\rangle$.

\begin{figure}
	\centering
	\includegraphics[width=\linewidth]{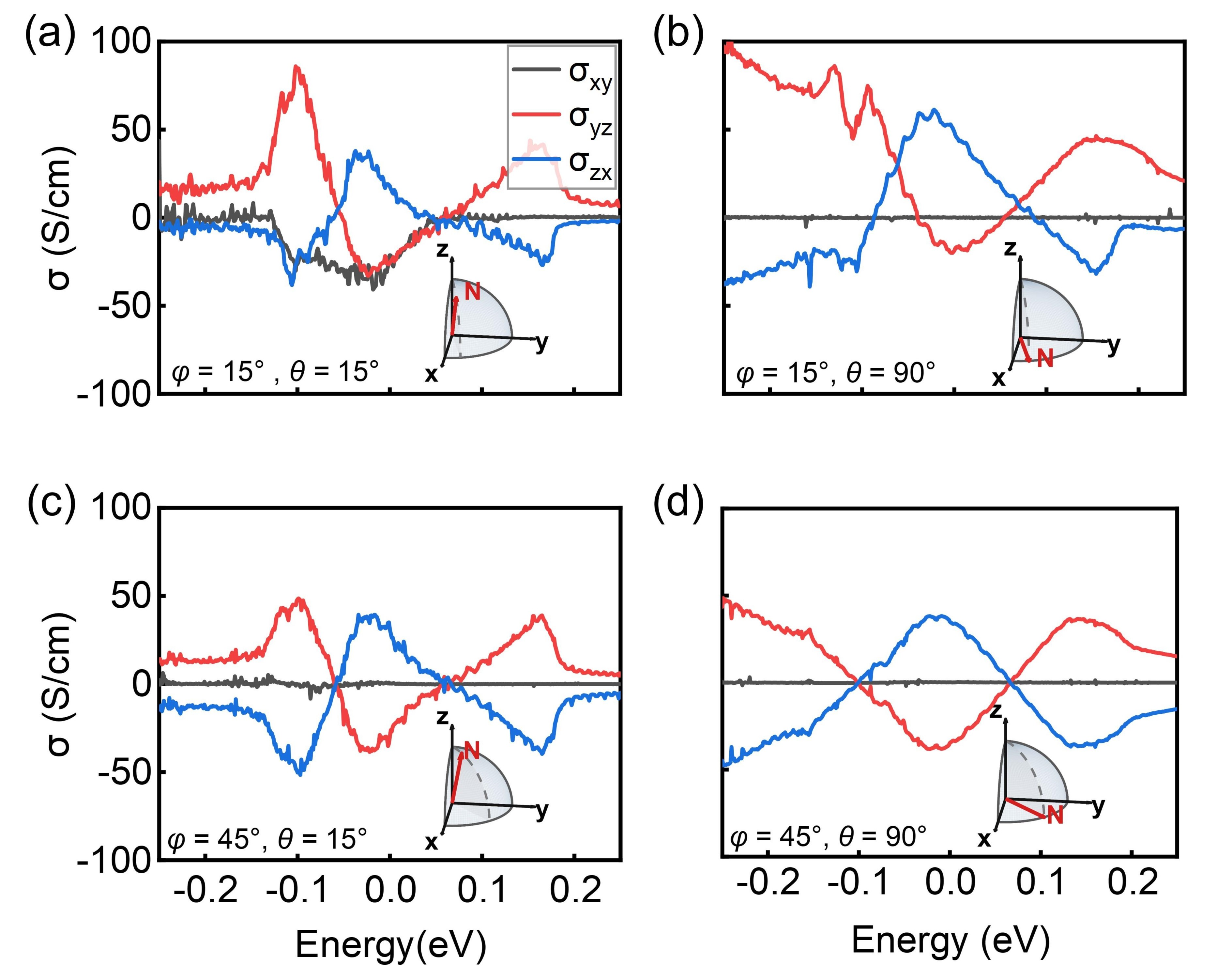}
	\caption{\textbf{Anomalous Hall conductivity for selected N\'eel-vector orientations in $\mathrm{V_2Te_2O}$.}
		(a,b) and (c,d) Energy-dependent anomalous Hall-conductivity
		components for $\varphi=15^\circ$ and $45^\circ$,
		respectively, with $\theta=15^\circ$ and $90^\circ$ shown in each pair.}
	\label{fig:fig5}
\end{figure}

This orbital mixing leaves clear signatures in the Berry-curvature distributions [Fig.~\ref{fig:fig3}(a-c)]. Under the residual $Im'm'm$ symmetry, $\Omega_{zx}$ and $\Omega_{xy}$ are either symmetry-forbidden or exhibit mutually cancelling pockets. In contrast, $\Omega_{yz}$ forms intense hotspots along the SOC-gapped nodal chain near the $X-M$ boundary. Band-resolved analysis along the representative $Q_1$--$P_1$--$Q_2$ path [Fig.~\ref{fig:fig3}(d-f)] confirms that bands undergoing an avoided crossing near the $E_F$ (gap at $P_1$) accumulate an enormous positive $\Omega_{yz}$ weight.

Crucially, this giant cross-layer Hall response is driven primarily by band anticrossings rather than isolated Weyl nodes. Although a pair of isolated Weyl nodes coexists along $\Gamma-X$ at $(\pm0.292,0,0)$ for $\theta=90^{\circ}$ (see Supplemental Materials S3 for detail), they yield no significant Berry-curvature weight at the AHC peak energy. Even at $\theta=45^{\circ}$, the Berry curvature remains overwhelmingly concentrated near the $P_1$ anticrossing region (more momentum-resolved Berry curvature distributions shown in Supplemental Materials S5).

\textit{Azimuthal Angle Control and Multi-Axial Switching.---} 
At $\varphi=0^\circ$ [Fig.~\ref{fig:fig4}(d)], the AHC obeys $\sigma_{yz}(\pi-\theta)=\sigma_{yz}(\theta)$, $\sigma_{xy}(\pi-\theta)=-\sigma_{xy}(\theta)$, and reverses sign upon a $180^\circ$ reversal of the N\'eel vector, $\sigma_{ij}(\theta+\pi)=-\sigma_{ij}(\theta)$.

Varying the azimuthal angle $\varphi$ enables multi-axial switching by redefining the residual magnetic symmetry [Fig.~\ref{fig:fig5}]. For low-symmetry orientation ($\varphi=15^\circ$, $\theta=15^\circ$), symmetry lowers to $P\bar{1}$, unlocking all three AHC components. At
$\varphi=15^\circ$ and $\theta=90^\circ$, restored $C_{2z}T$ and $M_zT$ symmetries forbid $\sigma_{xy}$, leaving independent $\sigma_{yz}$ and $\sigma_{zx}$ channels. In contrast, along the high-symmetry plane $\varphi=45^\circ$, residual diagonal operations $C_{2[1\bar{1}0]}$ and $M_{[1\bar{1}0]}$ impose $\sigma_{xy}=0$ and $\sigma_{yz}=-\sigma_{zx}$, leaving only one independent Hall-conductivity component. 
In short, polar angle ($\theta$) tunes the magnitude and sign of the AHC, while azimuthal angle ($\varphi$) dictates its tensor structure and channel multiplicity.

\textit{Conclusion.---}
In summary, we demonstrate that N\'eel-vector rotation in bulk V$_2$Te$_2$O enables symmetry-controlled activation and suppression of anomalous Hall transport. Tilting the N\'eel vector breaks the $M_z$ mirror symmetry, allowing SOC to gap the nodal chain and generate intense Berry-curvature hotspots through the enhancement of orbital hybridization across the layers. Crucially, the giant cross-layer Hall response is governed by these gapped anticrossing regions rather than isolated Weyl nodes, establishing a mechanism beyond the conventional Weyl-dominated picture.

Furthermore, the steep energy derivative of the AHC near the Fermi level suggests tensor-selective anomalous Nernst~\cite{GuoWang2017} and thermal Hall responses~\cite{Li2017} via the Mott relation. Our findings highlight this system as a promising platform for multi-axial spintronic and caloritronic devices.

\begin{acknowledgments}
This work was supported by the National Natural Science Foundation of China (Grants No.12595331), the National Key Research and Development Program of the Ministry of Science and Technology of China (No.2022YFA1602601),the National Natural Science Foundation of China (No.12574048, 12574257 and 12304214). The DFT calculations in this work are supported by the Beijing Super Cloud Computing Center (BSCC). 
\end{acknowledgments}

\bibliography{references}

\end{document}